\documentclass{article}
\usepackage{graphicx}

\usepackage{hyperref}
\usepackage{amssymb}
\usepackage{amsmath}
\usepackage{url}
\usepackage{amsbsy}
\usepackage{amsfonts}
\usepackage{caption}
\numberwithin{equation}{section}
\usepackage{amsthm}

\usepackage{graphicx}
\usepackage{amssymb}
\usepackage{url}
\usepackage[utf8]{inputenc}
\usepackage[english]{babel}
\usepackage{fancyhdr}
\usepackage{enumerate}
\usepackage{amssymb}
\usepackage{amsfonts}
\usepackage{amsmath}

\begin{document}

\date{
	\today
}

\title{On The Estimation Of Mutual Information}

\author{Nicholas Carrara $^{1}$, Jesse Ernst $^{2}$}



\maketitle



\abstract{In this paper we focus on the estimation of mutual information from finite samples $(\mathcal{X}\times\mathcal{Y})$.  The main concern with estimations of mutual information is their robustness under the class of transformations for which it remains invariant: i.e. type I (coordinate transformations), III (marginalizations) and special cases of type IV (embeddings, products).  Estimators which fail to meet these standards are not \textit{robust} in their general applicability.  Since most machine learning tasks employ transformations which belong to the classes referenced in part I, the mutual information can tell us which transformations are most optimal\cite{Carrara_Ernst}.  There are several classes of estimation methods in the literature, such as non-parametric estimators like the one developed by Kraskov et. al\cite{KSG}, and its improved versions\cite{LNC}.  These estimators are extremely useful, since they rely only on the geometry of the underlying sample, and circumvent estimating the probability distribution itself.  We explore the robustness of this family of estimators in the context of our design criteria.}

\section{Introduction}
Interpretting mutual information (MI) as a measure of correlation has gained considerable attention over the past couple of decades for it's application in both machine learning \cite{Carrara_Ernst,InfoBottle,InfoBottle2,InfoGAN} and in dimensionality reduction \cite{Sieve}, although it has a rich history in Communication Theory, especially in applications of Rate-Distortion theory \cite{CoverThomas}\footnote{While (MI) has been exploited in these examples, it has only recently been derived from first principles\cite{Carrara_MI}.}.  (MI) has several useful properties, such as having a lower bound of $I_{\min} = 0$ for variables which are uncorrelated.  Mostly we are interested in how the (MI) changes whenever we pass one set of variables through a function, which according to the data processing inequality can only ever destroy correlations and not increase them.  Thus, the (MI) for any set of variables $\mathbf{X}\times\mathbf{Y}$ is an upper bound for any transformation $f(\mathbf{X})\times g(\mathbf{Y})$.  This feature makes (MI) a good measure of performance for any (ML) task, which is why has gained so much attention recently \cite{Carrara_Ernst,InfoBottle,InfoBottle2,InfoGAN}.\\
\indent  The main challenge with using (MI) in any inference task is computing it when one only has a sample $\mathcal{X}\times\mathcal{Y} \subset \mathbf{X}\times\mathbf{Y}$.  The (MI) one estimates from the sample is highly dependent on the assumptions about the underlying joint distribution $p(x,y)$; effectively, one estimates (MI) by estimating the density $p(x,y)$.  The most popular method for estimating (MI) is by using the class of non-parametric estimators built on the method derived by Kraskov. et. al. \cite{KSG} (KSG).  The (KSG) estimator uses local geometric information about the sample to approximate the density $p(x_i,y_i)$ at each point $(x_i,y_i)$ and then calculates a local estimate of (MI) from it.  While this approach has been very successful, there are some weaknesses which we will discuss in this paper.  Specifically, the use of local geometric information causes the estimator to not be coordinate invariant in general, which is a violation of the basic properties of (MI).  What's worse, is it's inability to see through useless information, i.e. noise.  This is also a consequence of using local information without regard to the overall global structure of the space.  When combined, these two problems cause unwanted behavior in even the simplest of situations.\\
\indent  There have been some suggested improvements to (KSG) \cite{LNC} which we will discuss.  Most often studies of estimators are concerned with its effectiveness with small numbers of samples in large dimension, while here we will be mostly concerned with its robustness under coordinate transformations, redundancy and noise.  We will define redundancy and noise more precisely in a later section, but one can also check \cite{Carrara_MI} for a more rigorous definition.  As was shown in \cite{Carrara_Ernst} (KSG) handles redundant information well, which we will reiterate in a later section.  It is (KSG)'s inability to handle noise that diminishes it's effectiveness in real data sets.  In the next section we will briefly discuss the basic properties of (MI), and then discuss the ideas behind non-parametric entropy estimators.  We will then examine the robustness of (KSG) and it's improvements in section IV.  We end with a discussion.

\section{Mutual Information}
We will reiterate some the basic properties of mutual information.  Consider two spaces of propositions, $\mathbf{X}$ and $\mathbf{Y}$, whose joint space is given by $\mathbf{X}\times\mathbf{Y}$\footnote{The spaces $\mathbf{X},\mathbf{Y}$ can be either discrete/categorical or continuous.}.  The \textit{global} correlations present between the two spaces is determined from the mutual information,
\begin{equation}
I[\mathbf{X};\mathbf{Y}] = \int dxdy\, p(x,y)\log\frac{p(x,y)}{p(x)p(y)}\label{MI},
\end{equation}  
where $p(x,y)$ is the joint probability density and,
\begin{equation}
p(x) = \int dy\, p(x,y) \qquad \mathrm{and} \qquad p(y) = \int dx\, p(x,y),
\end{equation}
are the marginals.  The product marginal $p(x)p(y)$ can be interpreted as an independent prior and the (MI) gives a \textit{ranking} of joint distributions $p(x,y)$ according to their \textit{amount} of correlation; joint distributions with \textit{more} correlation have higher values of (MI).  The (MI) is bounded from below by $I_{\mathrm{min}} = 0$ whenever the spaces $\mathbf{X}$ and $\mathbf{Y}$ are uncorrelated; i.e. $p(x,y) = p(x)p(y)$, and is typically unbounded from above (except in cases of discrete distributions).\\
\indent  One immediate consequence of the functional form of (\ref{MI}) is its invariance under coordinate transformations.  Since the probabilities,
\begin{equation}
p(x,y)dxdy = p(x',y')dx'dy',
\end{equation}
are equivalent, then $I[\mathbf{X};\mathbf{Y}] = I[\mathbf{X}';\mathbf{Y}']$.  While this fact is somewhat trivial on its own, when combined with other types of transformations it can be quite powerful.  In this paper we will study three main types of transformations, the first being coordinate transformations.  The second kind of transformation of interest is \textit{marginalization}, and the third is \textit{products}.  Marginalization is simply the the projecting out of some variables, which according to the design criteria of (MI)\cite{Carrara_MI}, can only ever decrease the correlations present.  On the other hand, products of spaces can increase correlations when the new variables provide new information.  The most trivial type of product is an embedding, which is discussed in the next section.  
\subsection{Redundancy}
One advantage of (MI) is its invariance under the inclusion of redundant information.  For example, consider adding to the space $\mathbf{X}$ another space which is simply a function of $\mathbf{X}$, i.e. $\mathbf{X} \rightarrow \mathbf{X}\times f(\mathbf{X})$.  The joint probability distribution becomes,
\begin{equation}
p(x,f(x),y) = p(x,f(x))p(y|x,f(x)) = p(x)\delta(f(x) - f)p(y|x)\label{redundant},
\end{equation}
and hence the (MI) is,
\begin{equation}
I[\mathbf{X}\times f(\mathbf{X});\mathbf{Y}] = \int dxdydf\,p(x)\delta(f(x) - f)p(y|x)\log\frac{p(x)p(y|x)}{p(x)p(y)} = I[\mathbf{X};\mathbf{Y}]\label{red}.
\end{equation}
The map $\mathbf{X}\rightarrow \mathbf{X}\times f(\mathbf{X})$ is an embedding of $\mathbf{X}$ into a higher dimensional space.  Such a transformation does not increase the intrinsic dimension of $\mathbf{X}$.  Machine learning algorithms exploit this type of transformation in conjunction with coordinate transformations and marginalizations.

\subsection{Noise}  
Much like in eq. (\ref{redundant}), (MI) is also invariant under the addition of \textit{noise}, which are defined as variables, $\mathbf{Z}$, that are uncorrelated to both $\mathbf{X}$ and $\mathbf{Y}$,
\begin{equation}
p(x,z,y) = p(x,z)p(y|x,z) = p(x)p(z)p(y|x) = p(z)p(x,y).
\end{equation}
And like (\ref{red}) the mutual information is invariant,
\begin{equation}
I[\mathbf{X}\times\mathbf{Z};\mathbf{Y}] = I[\mathbf{X};\mathbf{Y}\times\mathbf{Z}] = \int dxdydz\,p(z)p(x,y)\log\frac{p(x,y)}{p(x)p(y)} = I[\mathbf{X};\mathbf{Y}].
\end{equation}
Unlike with redundancy, noise variables necessarily increase the dimension of the underlying space.

\section{Non-Parametric Estimation}
Non parametric entropy estimators attempt to utilize the geometry of the underlying sample to estimate the \textit{local} density and hence the \textit{local} entropy.  A popular estimator is the one developed by Kozachenko and Leonenko\cite{LK}, which we will briefly motivate.  Consider the task of estimating the entropy of a sample $\mathcal{X}$ from an underlying space $\mathbf{X}$.  Our goal is to find an unbiased estimator of the form $\hat{H}[\mathcal{X}] = N^{-1}\sum_{i=1}^N\log p(x_i)$, which converges to the \textit{true} Shannon entropy as $N\rightarrow\infty$\footnote{We highlight the word true here, since the underlying probability distribution is not known and hence our \textit{estimation} depends on our assumptions about its form.}.  To find an approximation of $\log p(x_i)$, consider the following probability distribution,
\begin{equation}
P_{\varepsilon}(x_i)d\varepsilon = \frac{(N_1)!}{1!(k-1)!(N - k - 1)!}p_i^{k-1}(1 - p_i)^{N-k-1}\frac{d p_i}{d\varepsilon}d\varepsilon,
\end{equation}    
which is the probability that the $k$th-nearest neighbor of the point $x_i$ exists within the small spherical shell of radius $\varepsilon/2$ and that there are $k-1$ points at $r_i < \varepsilon/2$ and $N-k-1$ points at $r_i > \varepsilon/2 + d\varepsilon$.  This distribution is of course properly normalized, and upon evaluating the expected value of the logarithm of $p_i$, we find,
\begin{equation}
\langle \log p_i\rangle = \int d\varepsilon\, P_\varepsilon(x_i)\log p_i = \psi(k) - \psi(N),
\end{equation}
where $\psi(k)$ is the digamma function.  From here one can determine an approximation for the logarithm of the \textit{true} distribution by assuming something about the local behavior of $p(x_i)$ with respect to the probability mass $p_i$.  In the (KL) approximation (and as well in the KSG approximation), it is assumed that the probability within the region defined by $p_i$ is uniform with respect to the \textit{true} distribution at the point $x_i$,
\begin{equation}
p_i \approx c_d \varepsilon^d p(x_i)\label{uniform},
\end{equation}
where $d$ is the dimension of the space and $c_d$ is the volume of the unit $d$-ball\footnote{The form of $c_d$ depends on the choice of metric for the space $\mathbf{X}$.  As we will see a useful choice for (MI) estimation is the $L^{\infty}$ norm.}.  Putting (\ref{uniform}) into the unbiased estimator one arrives at,
\begin{equation}
\hat{H}[\mathcal{X}] = \psi(N) - \psi(k) + \log c_d + \frac{d}{N}\sum_{i=1}^N\log(\varepsilon_i)\label{entropy}.
\end{equation}

\subsection{The Vanilla KSG Estimator}
The (KSG) estimator of the first kind is derived by taking the expression in (\ref{entropy}) and applying it to the decomposition,
\begin{equation}
\hat{I}[\mathcal{X};\mathcal{Y}] = \hat{H}[\mathcal{X}] + \hat{H}[\mathcal{Y}] -\hat{H}[\mathcal{X},\mathcal{Y}],
\end{equation}
where $\hat{H}[\mathcal{X},\mathcal{Y}]$ is the entropy over the joint distribution $p(x,y)$.  As has been shown and argued by (KSG), the approximation above is slightly naive since the local densities in the joint and marginal spaces can be different, leading to errors in the terms involving $\log(\varepsilon_i)$ which don't necessarily cancel.  As a neat trick, (KSG) suggests using the same density found in the joint space in the marginal spaces, so that the factors $(d_x/N)\sum_{i=1}^N\log(\varepsilon^x_i)$,$(d_y/N)\sum_{i=1}^N\log(\varepsilon^y_i)$ and $((d_x+d_y)/N)\sum_{i=1}^N\log(\varepsilon_i^{xy})$ will cancel.  Choosing the same density for fixed $k$ in the joint space causes the $k$ values in the marginal spaces to vary, and hence we arrive at the expression,
\begin{equation}
\hat{I}^1[\mathcal{X};\mathcal{Y}] = \psi(k) + \psi(N) - \langle \psi(n_x + 1)\rangle - \langle \psi(n_y + 1)\rangle,
\end{equation}
where $n_x$ and $n_y$ are the number of points which land in the $d_x$ and $d_y$ balls of radius $\varepsilon/2$ in the marginal spaces.\\
\indent  One unfortunate consequence of the (KSG) estimator of the first kind is its reliance on the $L^{\infty}$ norm for finding neighbors.  As has been pointed out by others\cite{LNC}, such a choice can lead to problems in regions where the probability varies greatly, which can easily happen in spaces of large dimension.  Unless the density of samples increases exponentially with respect to the dimension of the space, the errors in choosing $L^{\infty}$ will compound quickly.\\
\indent  In an alternative derivation, (KSG) attempts to derive another estimator which uses different distances in each of the marginal directions for the region of assumed uniform probability, in a sense replacing the $L^{\infty}$ box by a rectangular box whose side lengths are determined by some criteria.  While the idea would likely correct the errors accompanying the $L^{\infty}$ box, (KSG) were unsuccessful in deriving a closed form expression for their estimator of the second kind, $\hat{I}^2[\mathcal{X};\mathcal{Y}]$.  

\subsection{The LNC Correction to KSG}
As an attempted correction to (KSG)'s problem with using the $L^{\infty}$ box, S. Gao et. al. proposed the \textit{local non-uniform correction} technique.  This technique adjusts the unbiased estimator for (MI) by replacing the $L^{\infty}$ volume in the joint space with a volume computed from a PCA analysis.  The basic idea is the following.  Consider a point $x_i$ whose $k$th-neighbor is $x_k$.  With the collection of $k+1$ points including  $x_i, x_k$ and all points closer than $x_k$, construct the correlation matrix $C_{ij}$ and find its eigenvectors.  By then projecting each point along the maximal eigenvectors, we can find a PCA bounding box, which is rotated and skewed with respect to the $L^{\infty}$ box.  The assumption in this case is that the rotated PCA box is a much better representation of the region of uniform probability around $x_i$.  Once each volume is found, the (MI) is given by,
\begin{equation}
\hat{I}_{LNC} = \hat{I}_{KSG} - \frac{1}{N}\sum_{i=1}^N\log\frac{\bar{V}_i}{V_i}\label{LNC_1},
\end{equation}
where $\bar{V}_i$ is the PCA volume and $V_i$ is the $L^{\infty}$ volume.  Such an estimator has shown to give vast improvement to the naive KSG method, however current results are limited to two dimensional problems.  The reason for this is its inability to deal with redundant information.  To see this, consider a two-dimensional problem in which the variables $\mathbf{X}\times\mathbf{Y}$ have some non-trivial correlations.  If we add to $\mathbf{X}$ a redundant copy, $\mathbf{X}\rightarrow\mathbf{X}\times f(\mathbf{X})$, then we expect the (MI) to be invariant.  If one naively uses the (LNC) method, one will find that the (MI) increases.  This is because the volumes $\bar{V}_i$ will most often decrease when computed in the redundant scenario and hence the (LNC) correction term will most often increase.\\
\indent  A possible fix to this problem is to not only correct the volume in the joint space, but to fix the volumes in the marginal spaces as well, leading to the (LNC) correction of the second kind,
\begin{equation}
\hat{I}_{LNC^2} = \hat{I}_{KSG} - \frac{1}{N}\sum_{i=1}^N\log\frac{\bar{V}^{xy}_i}{\bar{V}^x_i\bar{V}^y_i}\label{LNC_2}.
\end{equation}
In investigating the efficacy of such a method, we discovered that it's not very robust.  This is mainly due to the fact that like the original method in (\ref{LNC_1}), (\ref{LNC_2}) is not coordinate invariant in general, and while the volume supplied by redundant variables can in principle be canceled in the denominator of (\ref{LNC_2}), the volumes themselves will be computed with respect to spaces of different dimension, and will therefore not exactly cancel.  The effect is to still increase (MI) under the influence of redundant variables, which is undesirable since vanilla (KSG) is most successful in this domain.  Thus while (LNC) fixes one aspect of (KSG), it reduces its efficacy in another aspect.

\section{Robustness Tests of NP Estimators}
We will access the robustness of the (KSG) estimator and its variants with respect to the three types of transformations outlined in section two (coordinate transformations, redundancy and noise).  Most tests in this section will use a multivariate normal distribution, 
\begin{equation}
\mathcal{N}_k = ((2\pi)^k|\mathbf{\Sigma}|)^{-1/2}\exp\left[-\frac{1}{2}(\mathbf{x} - \mu)^T\mathbf{\Sigma}^{-1}(\mathbf{x} - \mu)\right],
\end{equation}
where $\mathbf{\Sigma}$ is the covariance matrix. The mutual information between two sets of variables $\mathbf{X}_n$ and $\mathbf{X}_m$, where $n+m=k$, is given by,
\begin{equation}
I_{\mathcal{N}_k}[\mathbf{X}_n;\mathbf{X}_m] = -\frac{1}{2}\log\left(\frac{|\Sigma_k|}{|\Sigma_n||\Sigma|_m|}\right),
\end{equation}    
where $|\mathbf{\Sigma}_k|$ is the determinant of the covariance matrix $\mathbf{\Sigma}_k$.  For computing the (KSG) estimate, we will use a python package called (NPEET)\cite{npeet} developed by G. ver Steeg et. al.  For the (LNC) correction we use a similar package \cite{LNC}.  

\subsection{Coordinate Transformations}
Since (KSG) uses the $L^{\infty}$ norm to define the region of uniform probability for the estimate of $p_i$, this automatically presents a problem with coordinate invariance.  (KSG) will not even be invariant under linear scalings of the data, let alone arbitrary coordinate transformations\footnote{This is likely the motivation for (KSG)'s estimator of the second kind, which gave different weight to each of the variables, however they were unable to derive a closed form expression.}.  Essentially, if one variable, say $z$, is scaled by a large order of magnitude with respect to the other variables, then the side lengths of the $L^{\infty}$ box will get chosen to be the length of the $k$th nearest neighbor in the direction of $z$.  While this will not necessarily cause a problem with the values of $\langle \psi(x_z + 1)\rangle$, it will cause the counts for the other variables to be much larger than they necessarily should be.  As an example, consider the following bivariate normal case,
\begin{figure}[!h]
	\centering
	\includegraphics[width=.49\linewidth]{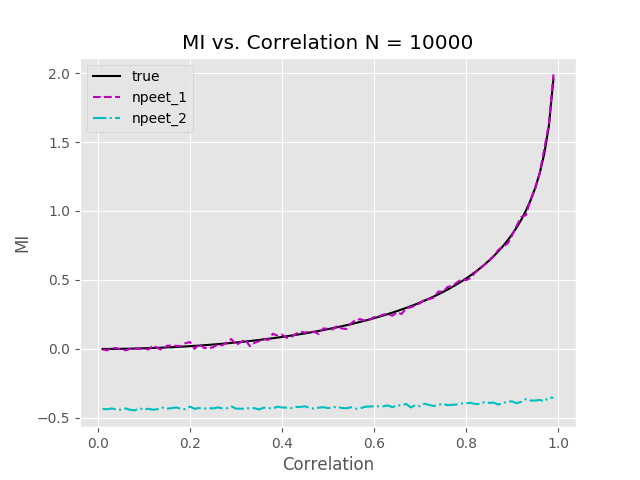}\includegraphics[width=.49\linewidth]{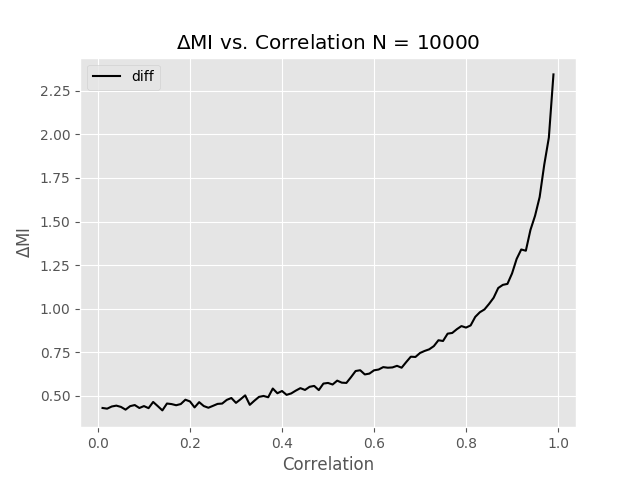}
	\caption{Comparison of mutual information estimates (KSG) for a bivariate normal distribution before (npeet\_1) and after (npeet\_2) a linear transformation of one variable by a factor of $10^5$.  The second plot shows the difference between (npeet\_1) and (npeet\_2).}
\end{figure}
As one can see from the figure, scaling one variable of the bivariate normal by $10^5$ renders the (KSG) method useless.  While one can always argue for a heuristic scaling method, any robust method for computing (MI) should be invariant under arbitrary scalings.  Below is the same experiment using the (LNC) method,
\begin{figure}[!h]
	\centering
	\includegraphics[width=.49\linewidth]{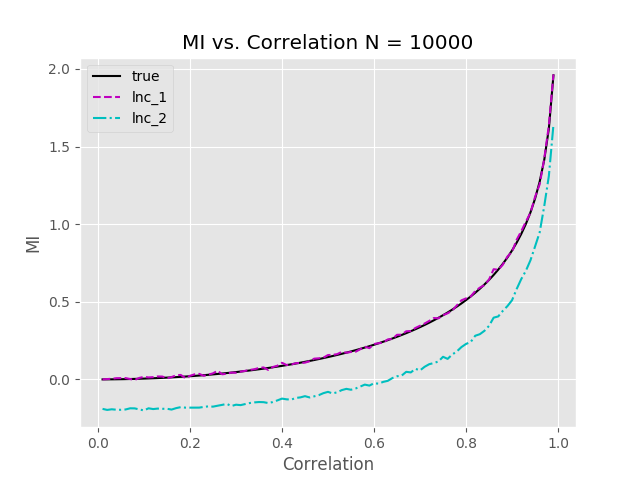}\includegraphics[width=.49\linewidth]{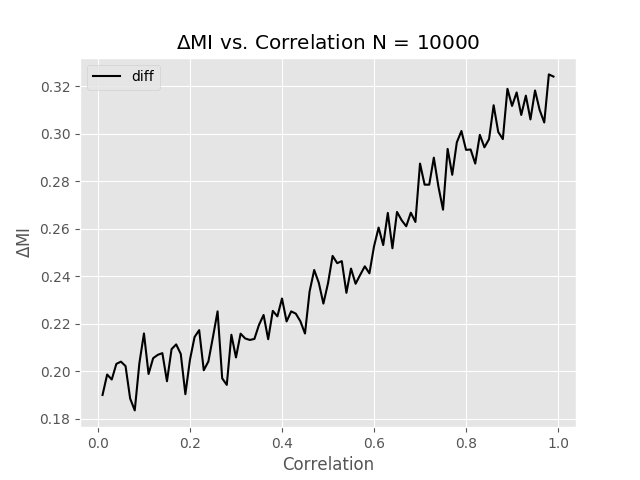}
	\caption{Comparison of mutual information estimates (LNC) for a bivariate normal distribution before (lnc\_1) and after (lnc\_2) a linear transformation of one variable by a factor of $10^5$.  The second plot shows the difference between (lnc\_1) and (lnc\_2).}
\end{figure}
While the (LNC) method corrects the behavior of the curve in the highly correlated region, it still fails to capture the correct value overall due to (KSG)'s inability to handle arbitrary coordinate transformations.  The effects of arbitrary coordinate transformations are even worse in higher dimensional situations.  Consider the eight-dimensional multivariate Gaussian below,
\begin{figure}[!h]
	\centering
	\includegraphics[width=.49\linewidth]{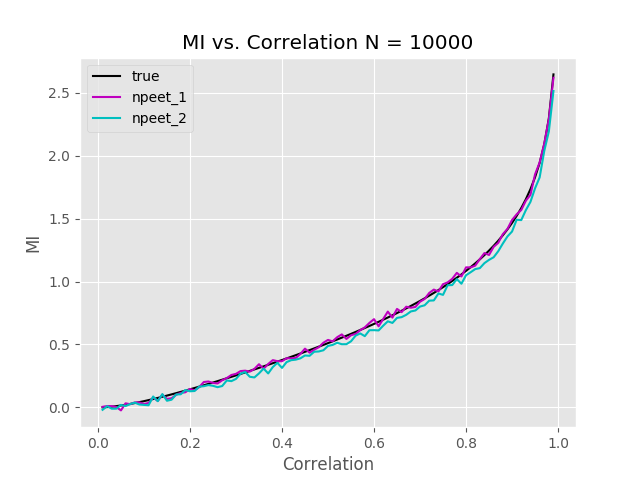}\includegraphics[width=.49\linewidth]{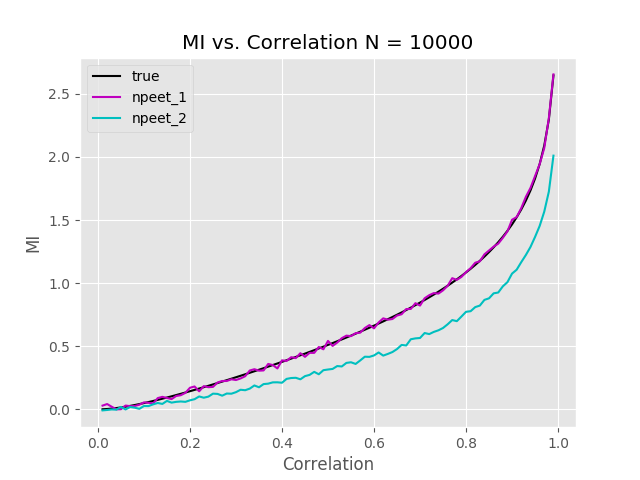}
	\caption{Comparison of mutual information estimates (KSG) for a multivariate normal distribution with equal correlation coefficients $\rho_{ij}$ before (npeet\_1) and after (npeet\_2) a linear transformation of one variable by a factor of $10$.  The second plot compares the (KSG) estimators after four variables are multiplied by a factor of $10$.}
\end{figure}
As you can see, multiplying all the variables by a factor of 10 greatly reduces the accuracy of the KSG algorithm.  To see this effect happening more gradually, we focus on one particular value of the correlation matrix (where all correlation coefficients are equal to $1/2$) and dial up the transformation on the four variables in one set.  The results for 10,000 and 100,000 points are below,
\begin{figure}[!h]
	\centering
	\includegraphics[width=.49\linewidth]{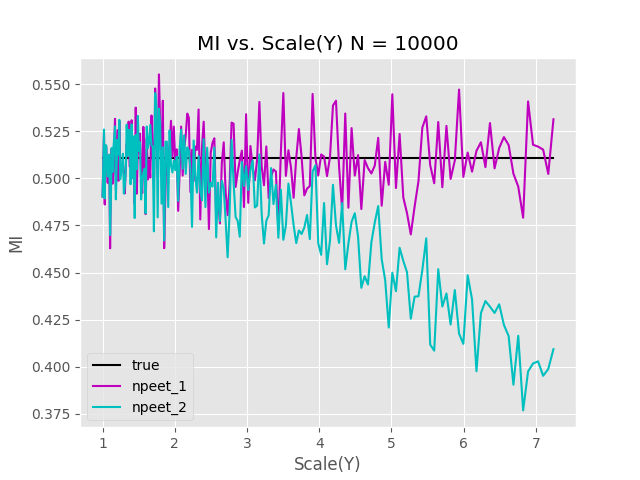}\includegraphics[width=.49\linewidth]{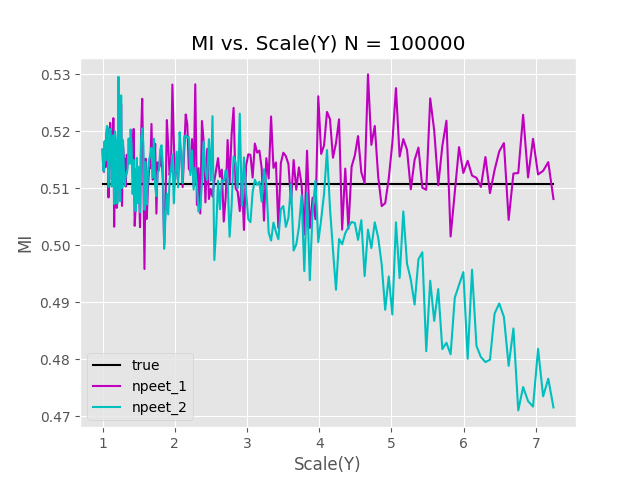}
	\caption{Comparison of true mutual information between the eight variables of a multivariate normal distribution where all correlation coefficients are equal to $1/2$ (black line) and the value estimated from $N=10,000$ and $N=100,000$ samples before (npeet\_1) and after (npeet\_2) a linear transformation is applied to one variable $x \rightarrow x' = 10x$.}
\end{figure}
As you can see, the KSG approximation begins to deteriorate very quickly under linear transformations when the dimensionality is high.  Increasing the number of points by a factor of ten seems to do little to help this.  While this is certainly a flaw in the method, it isn't as dire as the others we will explore in the next section.  For now, one can adopt a strategy in which each variable is scaled in a way that gives equal weight to each of them.  Proposed methods for this were given in \cite{Carrara_Ernst}.

\subsection{Redundancy vs. Noise}
Another simple test we can perform is to see how MI in high dimensional situations handles redundant and noisy variables.  Specifically, we will look at how the MI changes as we dial up the noise present in redundant variables by randomizing their values with respect to the other variables.  As we saw in \cite{Carrara_Ernst}, (KSG) does quite well with redundant variables, however noisy variables still present a problem.\\
\indent  We have studied the ability of vanilla (KSG) to calculate (MI) under the presence of redundant variables in \cite{Carrara_Ernst}.  Here we will briefly discuss the highlights.  We examined a binary decision problem in which two distributions (signal and background) are separated by a certain amount with respect to their means.  In particular we study five-dimensional Gaussians with means $\mu_s = 1$, $\mu_n = -1$ and variances $\sigma^2 = 1$ for both signal and background.  We calculate (MI) using (NPEET) on a sample of size $N=10,000$ and then compare that value to the (MI) of a neural network output on the same sample after the network was trained for 100 epochs to distinguish between the signal and background distributions.  We showed in \cite{Carrara_Ernst} that the neural network transformation leaves the (MI) unchanged which is expected according to its design criteria \cite{ME}.  In another example, we added redundant variables that were functions of the original five and observed that (KSG) did not increase the (MI) as expected with redundant variables.  The examples are shown in the plots below,
\begin{figure}[!h]
	\centering
	\includegraphics[width=.49\linewidth]{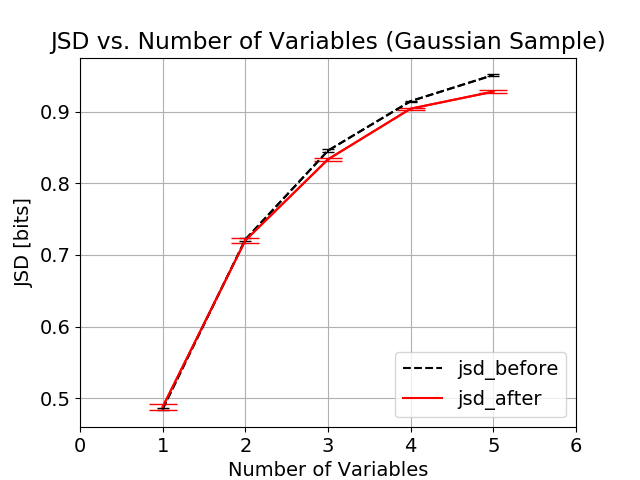}
	\includegraphics[width=.49\linewidth]{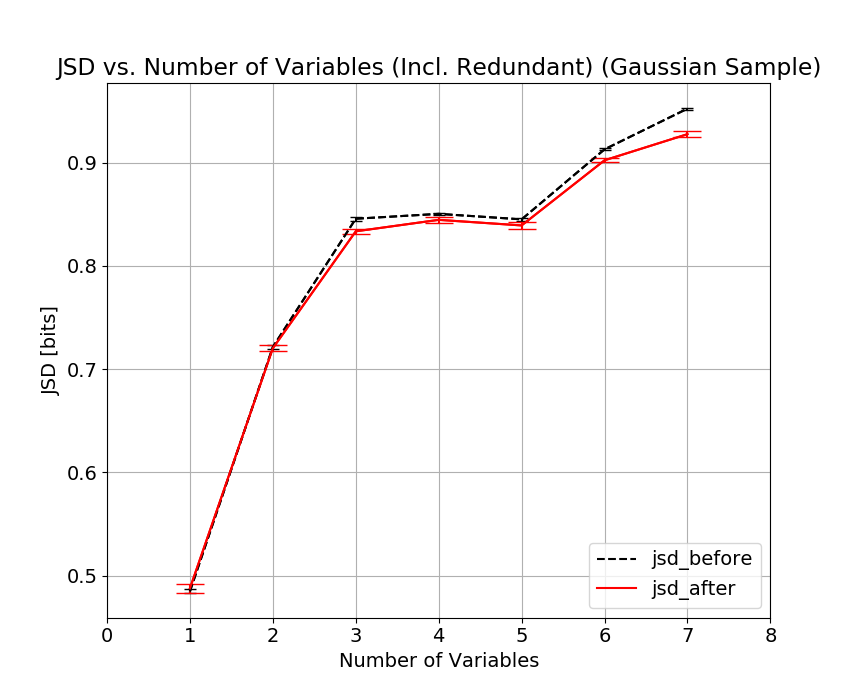}
	\caption{Comparison of (MI) values for increasing additions of discriminating variables for a five dimensional Gaussian.  The first plot compares a (NN) performance on the five dimensional Gaussian variables, while the second plot shows how (KSG) remains invariant after including redundant variables (three and four).}
\end{figure}
As we can see, (KSG) handles redundant information well.  We tested this claim on a more general data set which was generated as part of a machine learning analysis on a mock SUSY search \cite{Whiteson}.  The SUSY data set contains eight low-level variables and ten high-level variables which are functions of the low-level ones.  From the tests in \cite{Carrara_Ernst}, we again see that (KSG) performs well under the addition of the high-level variables when they are redundant,
\begin{figure}[!h]
	\centering
	\includegraphics[width=.49\linewidth]{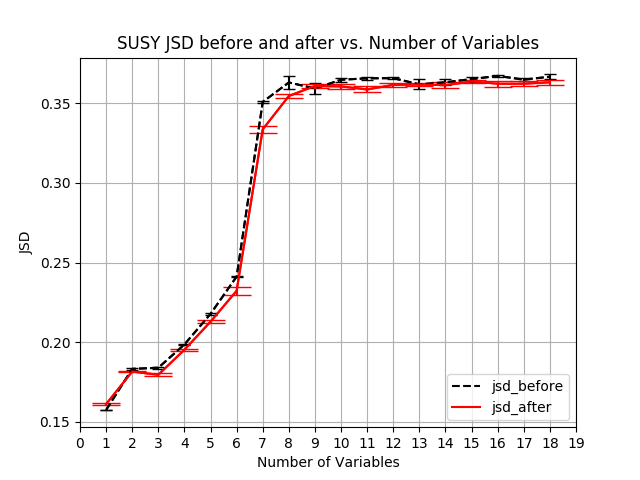}
	\includegraphics[width=.49\linewidth]{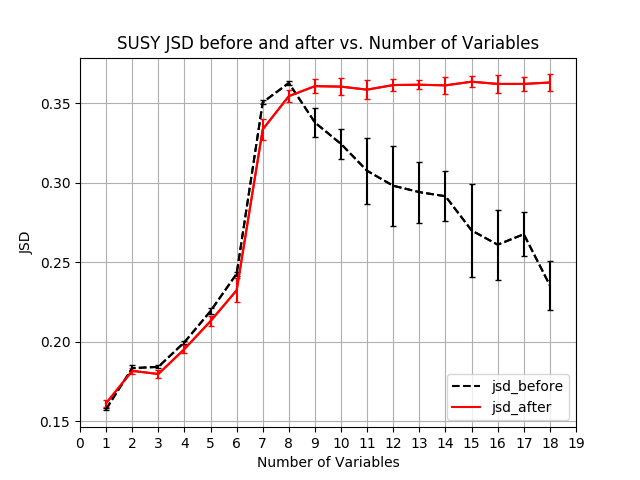}
	\caption{Comparison of (MI) values for increasing additions of discriminating variables for the SUSY data set.  The first eight variables are low-level and the last ten are functions of the first eight.  The first plot shows a comparison of (NN) performance when the last ten variables are redundant while the second shows how (KSG)'s accuracy deteriorates when the high-level variables are shuffled.}
\end{figure}
However when the variables are shuffled so that they are independent of both $\mathbf{X}$ and $\mathbf{\Theta}$, i.e. when they are noise, then (KSG) starts to deteriorate very quickly.  We can see this effect in the case where the (MI) is known exactly.  Using the multi-variate Gaussian with eight variables from the previous tests, we added three copies of the last variable on one side, essentially taking $x_4 \in X$ and creating $X' = X \cup x_4 \cup x_4 \cup x_4$.  We then dialed up the randomness in the three variables to 100\%.  What we mean here is that we rearranged the points in the set $(x_4\cup x_4\cup x_4)$ so that a particular value $x_i \in (x_4\cup x_4\cup x_4) \neq f(\mathbf{X})$ and $x_i \in (x_4\cup x_4\cup x_4) \neq f(\mathbf{Y})$,
\begin{figure}[!h]
	\centering
	\includegraphics[width=.49\linewidth]{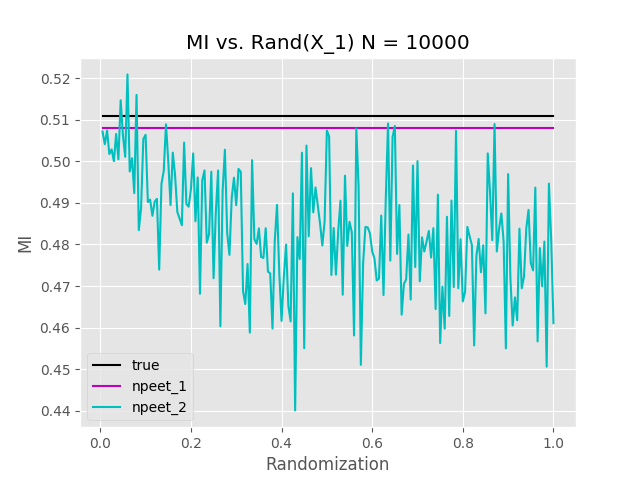}\includegraphics[width=.49\linewidth]{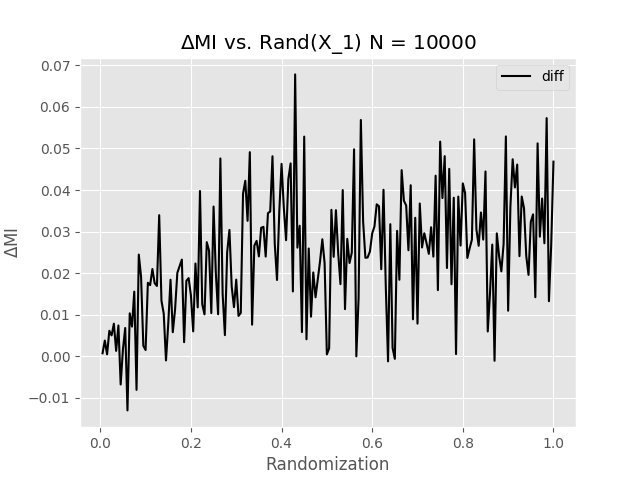}
	\caption{Comparison of true mutual information between the eight variables of a multivariate normal distribution where all correlation coefficients are equal to $1/2$ (black line) and the value estimated from $N=10,000$ samples before (npeet\_1) and after (npeet\_2) a randomization is applied to the set of redundant variables.}
\end{figure}
This shows that noisy variables cause the (KSG) estimate to go down as a function of their uselessness.  This problem compounds quickly when the number of useless variables increases, making (KSG)'s ability to determine (MI) in high-dimensional cases problematic.  

\section{Discussion}
We've shown in practical examples that (KSG) is not robust under coordinate transformations and noise.  While the effect of including noise is not as drastic in cases of simple distributions (figure 7), it is must more dramatic in cases where the the distribution is not simple (figure 6).

\section*{Acknowledgement}We would like to thank A. Caticha, K. Knuth, G. ver Steeg and K. Vanslette for insightful conversations.

\bibliography{maxentbib.bib}

\end{document}